\documentclass[12pt,letterpaper,english,aps,tightenlinesletterpaper,groupaddress,secnumarabic,nofootinbib]{revtex4}
\usepackage{mathptmx}

\usepackage[T1]{fontenc}
\usepackage[latin9]{inputenc}
\usepackage{color}
\usepackage{babel}
\usepackage{amsmath}
\usepackage{amssymb}
\usepackage{undertilde}
\usepackage{graphicx}
\usepackage{esint}
\usepackage[unicode=true,pdfusetitle,
 bookmarks=true,bookmarksnumbered=false,bookmarksopen=false,
 breaklinks=false,pdfborder={0 0 1},backref=section,colorlinks=false]
 {hyperref}
\hypersetup{
 citecolor=blue}

\makeatletter

\pdfpageheight\paperheight
\pdfpagewidth\paperwidth

\@ifundefined{textcolor}{}
{%
 \definecolor{BLACK}{gray}{0}
 \definecolor{WHITE}{gray}{1}
 \definecolor{RED}{rgb}{1,0,0}
 \definecolor{GREEN}{rgb}{0,1,0}
 \definecolor{BLUE}{rgb}{0,0,1}
 \definecolor{CYAN}{cmyk}{1,0,0,0}
 \definecolor{MAGENTA}{cmyk}{0,1,0,0}
 \definecolor{YELLOW}{cmyk}{0,0,1,0}
}

\usepackage{babel}

\usepackage{amsfonts}
\usepackage{bbm}
\usepackage{euscript}
\usepackage{dcolumn}
\usepackage{bm}
\usepackage{epsfig}\epsfclipon
\usepackage{slashed}

\usepackage{color}

\makeatother

\begin{document}

\title{Ward Identity and Basis Tensor Gauge Theory at One Loop}

\author{Daniel J. H. Chung}

\email{danielchung@wisc.edu}

\affiliation{Department of Physics, University of Wisconsin-Madison, Madison,
WI 53706, USA}
\begin{abstract}
Basis tensor gauge theory (BTGT) is a reformulation of ordinary gauge
theory that is an analog of the vierbein formulation of gravity and
is related to the Wilson line formulation. To match ordinary gauge
theories coupled to matter, the BTGT formalism requires a continuous
symmetry that we call the BTGT symmetry in addition to the ordinary
gauge symmetry. After classically interpreting the BTGT symmetry,
we construct using the BTGT formalism the Ward identities associated
with the BTGT symmetry and the ordinary gauge symmetry. As a way of
testing the quantum stability and the consistency of the Ward identities
with a known regularization method, we explicitly renormalize the
scalar QED at one-loop using dimensional regularization using the
BTGT formalism.
\end{abstract}
\maketitle

\section{Introduction}

Gauge theories (see e.g. \cite{Weyl:1919fi,Weyl:1929fm,Yang:1954ek,Abers:1973qs,Itzykson:1980rh,Polyakov:1987ez,'tHooft:1995gh,Weinberg:1996kr})
used to write Standard Model (SM) of particle physics \cite{Glashow:1961tr,Weinberg:1967tq,Salam:1968rm,Gross:1973id,Politzer:1973fx,Weinberg:1996kr,Ramond:1999vh,Langacker:2010zza,Aad:2012tfa,Chatrchyan:2012xdj}
are usually written in terms of fields which transform inhomogeneously
under the gauge group: i.e.~these are connections on principal bundles
(see e.g.~\cite{Nakahara:2003nw,Wu:1975es}). In \cite{Chung:2016lhv},
we constructed a reformulation of ordinary gauge theories in analogy
with the vierbein reformulation of general relativity. In particular,
we constructed a replacement for the gauge field degree of freedom
which transforms homogeneously under the $U(1)$ gauge group and satisfies
certain constraints. Since the lowest rank Lorentz tensor for such
a field was shown to be two, the field $G_{\,\,\,\,\,\beta}^{\alpha}$
(which replaces the usual $A_{\mu}$ gauge field) carries two Lorentz
indices, transforms as a multiplicative $U(1)$ phase representation,
and satisfies a non-linear constraint equation. 

This constraint equation was solved in \cite{Chung:2016lhv} in terms
of $N$ unconstrained scalars $\theta^{a}(x)$ ($N$ is the number
of spacetime dimensions) with the help of constant matrices $(H^{a})_{\,\,\,\,\,\,\nu}^{\mu}$
which in some sense are analogous to the Clifford algebra basis matrices
in the well known spinor field theory constructions. The field theory
of $\theta^{a}(x)$ is what we called \emph{basis tensor gauge theory}
(BTGT). The closest cousin of the $\theta^{a}$ field is the Wilson
line (e.g.~\cite{Wilson:1974sk,Giles:1981ej,Migdal:1984gj,Terning:1991yt,Gross:2000ba,Kapustin:2005py,Cherednikov:2008ua,Mandelstam:1962mi}
and references therein), but $\theta^{a}(x)$ has an appealing simplicity
of being manifestly local. 

To match the BTGT action to that of the usual gauge theory coupled
to scalars, we had to impose a \emph{new local symmetry} in \cite{Chung:2016lhv}
which we will call the BTGT symmetry. In this paper, we investigate
the Ward identities associated with the BTGT symmetry and the usual
$U(1)$ gauge symmetry within the BTGT formalism as a step in building
practical computational tools and checking the theory's quantum stability.
First, we find that the BTGT symmetry current itself can be classically
interpreted as a decomposition of $A^{\mu}$ equations of motion in
basis tensor components. We will also find a relationship of a particular
combination of BTGT current conservation and the residual gauge symmetry
current conservation in $\xi$-fixed ordinary $A^{\mu}$ field theory
formalism. Next, we use the effective action formalism to derive the
BTGT and the $U(1)$ Ward identities in both configuration space and
momentum space within the BTGT formalism. These identities are then
explicitly applied to 1-loop renormalization of scalar QED. We find
that dimensional regularization preserves the BTGT symmetry (in addition
to the $U(1)$ as expected). The explicit computations also highlight
the utility of the basis tensor star product $J\star_{c}K\equiv J^{\mu}(H^{c})_{\mu\nu}K^{\nu}$
in computing the Feynman graphs within the BTGT formalism. Through
this explicit renormalization exercise, we confirm that BTGT is stable
at one loop.

The order of presentation is as follows. In the next section, we give
a quick overview of the basis tensor gauge theory formalism. In Sec.~\ref{sec:Classical-BTGT-symmetry},
we construct the BTGT Noether current and give a classical interpretation.
In Sec.~\ref{sec:A-brief-review}, we briefly review the effective
action method of generating Ward identities. In Sec.~\ref{sec:BTGT-formalism-Ward},
we compute the Ward identities for both the BTGT symmetry and $U(1)$
symmetry in the context of the BTGT formalism. In Sec.~\ref{sec:1-loop-renormalization},
we use the BTGT formalism Feynman rules to explicitly renormalize
the scalar QED at 1-loop, checking the quantum stability of the theory
as well as the consistency of the dimensional regularization with
the Ward identities. We conclude in Sec.~\ref{sec:Conclusions} by
speculating on future research directions. The Appendix presents basic
identities of $(H^{a})_{\,\,\,\,\,\nu}^{\mu}$ useful for Feynman
diagram computations. There, we also point out a minor typo in equation
36 of \cite{Chung:2016lhv}

\section{Review of BTGT formalism}

In this section we give a very brief review of \cite{Chung:2016lhv}.
For more details, we refer the reader to the original article.

A vierbein formulation of Einstein gravity relies on finding a basis
of spacetime vector fields that transforms as an $(1,1)$ under $SO(3,1)\otimes{\rm diffeomorphism}$.
By contracting with vierbeins, a non-singlet diffeomorphism tensor
turns into a set of diffeomorphism scalar fields that transform as
a non-singlet $SO(3,1)$ tensor components. The vierbein's relationship
with the Christoffel symbols (i.e.~the gravitational gauge fields)
can be viewed as the vierbeins being solutions to a set of nonlinear partial
differential equations involving the Christoffel symbols. The vierbein
analog in the context in which ordinary compact Lie groups replace
diffeomorphism group is the focus of BTGT.

In the $U(1)$ BTGT of \cite{Chung:2016lhv}, the field $G_{\,\,\,\,\,\,\,\beta}^{\alpha}$
is the analog of the gravitational vierbein and it transforms as a
$(2,1)$ under $SO(3,1)\otimes U(1)$.  The constraint of matching
to the usual $U(1)$ connection (i.e.~the analog of the gravitational
vierbein relationship to the Christoffel symbols $\Gamma_{\mu\nu}^{\alpha}$)
is
\begin{equation}
A_{\mu}=-i(G^{-1})_{\,\,\,\,\,\,\,\beta}^{\alpha}\partial_{\alpha}G_{\,\,\,\,\,\,\,\,\mu}^{\beta}.
\end{equation}
The number of degrees of freedom in $G_{\,\,\,\,\,\,\,\beta}^{\alpha}$ is reduced by introducing BTGT fields $\theta^{a}$ (i.e.~4 scalar fields
in 4 spacetime dimensions) through
\begin{equation}
G_{\,\,\,\,\,\,\,\,\mu}^{\beta}(x)=\left(e^{i\theta^{a}(x)H^{a}}\right)_{\,\,\,\,\,\,\,\,\mu}^{\beta}
\end{equation}
where the constant matrices $H^a$ satisfy $[H^{a},H^{b}]=0$ and $H^{a}$ transforms like a rank 2-tensor
under Lorentz transformations.\footnote{This is analogous to normalizing the gravitational vierbein dot products to reduce to the Minkowski metric.} An explicit representation of $H^{a}$
is discussed in Sec.~\ref{sec:Useful-identities}. The relationship
of $\theta^{a}$ to $A_{\mu}$ is then
\begin{equation}
A_{\mu}(x)=\sum_{a}\left(H^{a}\right)_{\,\,\,\,\,\,\mu}^{\alpha}\partial_{\alpha}\theta^{a}
\end{equation}
which says that $\theta^{a}$ is similar to the Wilson line.

To construct a field theory of $\theta^{a}$ (i.e.~BTGT theory) that
matches the ordinary gauge theory, one must impose the following independent
continuous symmetries:
\begin{enumerate}
\item Ordinary gauge symmetry: if $\phi$ has charge $e$ under $U(1),$
then the $U(1)$ gauge symmetry is the set \{$\delta\phi=ie\theta\phi,\,\,\,\,\,\,\,\delta\phi^{*}=-ie\theta\phi^{\
*},\,\,\,\,\,\delta\theta^{a}=\theta$\}.
\item A BTGT symmetry: vary just the BTGT field by a restricted class of
functions $\delta\theta^{a}=\utilde{Z}^{a}(x)$ where $(H^{a})_{\,\,\,\,\,\,\mu}^{\alpha}\partial_{\alpha}\utilde{\
Z}^{a}(x)=0$
and there is no sum over the repeated $a$ index in this equation.
\end{enumerate}
At the renormalizable level, this set of symmetries reproduces the
ordinary $U(1)$ gauge theory action.  For example, in the case of scalar QED, the partition function is
\begin{equation}
\mathcal{Z}=\int  D\theta^a D\phi D\phi^{*}e^{i S[\theta^a,\phi,\phi^{*}]}
\end{equation}
where the Lagrangian terms for the $\xi$-gauge fixed action are given by Eq.~(\ref{eq:btgtL1})-(\ref{eq:BTGTL4}).

Correlators of certain differences
of $\theta^{a}$ map to correlators of integrals over $A_{\mu}$.

\section{\label{sec:Classical-BTGT-symmetry}Classical BTGT symmetry current}

To write scalar QED in the BTGT formalism, a new local symmetry (in
addition to the usual $U(1)$ gauge symmetry) was introduced in \cite{Chung:2016lhv}
which we will refer to as the \emph{BTGT symmetry} in this paper.
To have a physical interpretation of the continuous BTGT symmetry,
we construct the Noether current associated with this symmetry in
this section. More explicitly, we seek the Noether currents associated
with the BTGT symmetry defined as 
\begin{equation}
\delta\theta^{a}=\utilde{Z}^{a}(x)\label{eq:BTGTsymhere}
\end{equation}
satisfying the constraint equation
\begin{equation}
(H^{a})_{\,\,\,\,\,\,\mu}^{\alpha}\partial_{\alpha}\utilde{Z}^{a}(x)=0\label{eq:btgtgaugeconstraint}
\end{equation}
where there is no sum over the repeated $a$ index in this equation.
Note that even though $\utilde{Z}^{a}(x)$ is reminiscent of a pure
$U(1)$ gauge field function $\theta(x)$ appearing in $\delta A_{\mu}=\partial_{\mu}\theta$,
there is no gauge charged tensor transformation here since all the
gauge charged matter fields are held fixed and only the $\theta^{a}$
transforms as $\delta\theta^{a}=\utilde{Z}^{a}(x)$. Furthermore,
each $\utilde{Z}^{a}(x)$ for different $a$ indices are independent. 

To gain some intuition of the mathematical procedure for constructing
the Noether current of a constrained symmetry representation, it is
useful to review a mathematically analogous more familiar symmetry:
the residual gauge transformation of a $\xi$-gauge fixed scalar QED
theory. With the gauge fixing term
\begin{equation}
\mathcal{L}_{GF}=\frac{-1}{2\xi}(\partial A)^{2}\label{eq:gaugefix}
\end{equation}
there still exists a continuous residual gauge transformation
\begin{equation}
\delta A_{\mu}=\partial_{\mu}h,\,\,\,\,\,\delta\phi=ieh\phi,\,\,\,\,...\label{eq:transform}
\end{equation}
where 
\begin{equation}
\square h=0.\label{eq:residual-gauge-cond}
\end{equation}
Eq.~(\ref{eq:residual-gauge-cond}) is the analog of the constraint
Eq.~(\ref{eq:btgtgaugeconstraint}). Using the standard Noether construction
with the non-symmetry deformation parameterized through an arbitrary
continuous function $\epsilon(x)$ as%
\footnote{In principle, we can separate the functional variations more carefully
since some $\epsilon(x)$ choices will keep $\epsilon h$ harmonic.
Owing partly to Lorentz invariance, one can check that this does not
change the results for the construction of local currents. %
} 
\begin{equation}
\delta A_{\mu}=\partial_{\mu}[\epsilon(x)h(x)],
\end{equation}
it is straightforward to show
\begin{equation}
\partial_{\mu}j_{R}^{\mu}=0\label{eq:residual-gauge-current}
\end{equation}
where
\begin{equation}
j_{R}^{\mu}\equiv\partial^{\mu}(\partial_{\nu}A^{\nu}).
\end{equation}
Since the classical equations of motion in $\xi$-gauge satisfies
\begin{equation}
\partial_{\nu}\left[\partial_{\mu}F^{\mu\nu}+\frac{1}{\xi}\partial^{\nu}(\partial_{\mu}A^{\mu})-J_{U(1)}^{\nu}\right]=0\label{eq:residualcurrentconserve}
\end{equation}
where $J_{U(1)}^{\nu}$ is a Noether current associated with the global
subgroup of $U(1)$, we see that Eq.~(\ref{eq:residual-gauge-current})
indeed is satisfied owing to the antisymmetric property of the field
strength tensor and the global $U(1)$ current conservation. 

We can carry out the same exercise with the BTGT symmetry of Eq.~(\ref{eq:BTGTsymhere}).
We consider the $\xi$-gauge fixed scalar QED in the BTGT basis \cite{Chung:2016lhv}:
\begin{eqnarray}
\mathcal{L}_{k2} & = & \frac{-Z_{3}}{2}\partial_{\mu}\left(\partial_{\alpha}\theta^{a}(H^{a})_{\,\,\,\,\,\nu}^{\alpha}\right)\partial^{\mu}\left(\partial_{\beta}\theta^{b}(H^{b})^{\beta\nu}\right)\nonumber \\
 &  & +\frac{Z_{3}}{2}\partial_{\mu}\left(\partial_{\alpha}\theta^{a}(H^{a})_{\,\,\,\,\,\nu}^{\alpha}\right)\partial^{\nu}\left(\partial_{\beta}\theta^{b}(H^{b})^{\beta\mu}\right).\label{eq:btgtL1}
\end{eqnarray}
\begin{equation}
\mathcal{L}_{I}=Z_{1}\left[ie\phi^{*}\partial^{\mu}\phi\partial_{\alpha}\theta^{a}\left(H^{a}\right)_{\,\,\,\,\,\,\mu}^{\alpha}+h.c.\right]+Z_{4}e^{2}|\phi|^{2}\partial_{\alpha}\theta^{a}(H^{a})_{\,\,\,\,\,\mu}^{\alpha}\partial_{\beta}\theta^{b}(H^{b})^{\beta\mu}.\label{eq:btgtL2}
\end{equation}
\begin{equation}
\mathcal{L}_{ks}=Z_{2}|\partial\phi|^{2}-m^{2}Z_{m}|\phi|^{2}\label{eq:BTGTL3}
\end{equation}
\begin{equation}
\mathcal{L}_{{\rm GF}}\equiv\frac{-1}{2\xi}[\partial^{\mu}\partial_{\alpha}\theta^{a}(H^{a})_{\,\,\,\,\,\mu}^{\alpha}]^{2}.\label{eq:BTGTL4}
\end{equation}
In these expressions, the summation over repeated indices is implied,
and we will assume below that repeated indices are summed unless stated
otherwise explicitly or if it is clear from the context. This action
is invariant under the BTGT symmetry Eq.~(\ref{eq:BTGTsymhere}).
The gauge fixing term $\mathcal{L}_{GF}$ breaks the usual $U(1)$
gauge transformations which when written in the BTGT formalism are
\begin{equation}
\delta\phi=ie\theta\phi,\,\,\,\,\,\,\,\delta\phi^{*}=-ie\theta\phi^{*},\,\,\,\,\,\delta\theta^{a}=\theta\label{eq:ordinarygaugeu1inbtgt}
\end{equation}
where $\theta$ is an arbitrary smooth function, but it does preserve
the global $U(1)$ subgroup.

We find the conservation of Noether current associated with the BTGT
symmetry to be
\begin{equation}
\partial_{\mu}\mathcal{B}_{a}^{\mu}=0\label{eq:conservationbtgtcurrent}
\end{equation}
where
\begin{equation}
\boxed{\mathcal{B}_{a}^{\mu}=(H^{a})_{\,\,\,\,\,\lambda}^{\mu}\left[\partial^{\lambda}\sum_{b}\partial\star_{b}\partial\theta^{b}(1-\frac{1}{\xi})-\partial^{\lambda}\square\theta^{a}+\mathcal{C}_{U(1)a}^{\lambda}(x)\right]}\label{eq:btgtcurrent}
\end{equation}

\begin{equation}
\mathcal{C}_{U(1)a}^{\mu}\equiv-\left[ie\phi^{*}\partial^{\mu}\phi+h.c.\right]-2\left[e^{2}|\phi|^{2}\partial^{\mu}\theta^{a}\right]
\end{equation}
where we have introduced the notation 
\begin{equation}
\boxed{A\star_{b}B\equiv A^{\mu}B^{\nu}(H^{b})_{\mu\nu}}.
\end{equation}
This current is ordinary $U(1)$ gauge invariant if $\xi\rightarrow\infty$
since under Eq.~(\ref{eq:ordinarygaugeu1inbtgt}), the current variation
is 
\begin{eqnarray}
\delta\mathcal{B}_{a}^{\mu} & = & (H^{a})_{\,\,\,\,\,\lambda}^{\mu}\left[(1-\frac{1}{\xi})\partial^{\lambda}\left(\sum_{b}\partial\star_{b}\partial\right)\theta-\partial^{\lambda}\square\theta-\left[ie\phi^{*}\partial^{\mu}(ie\theta\phi)-ie\phi\partial^{\mu}(-ie\theta\phi^{*})\right]\right.\nonumber \\
 &  & \left.-2e^{2}|\phi|^{2}\partial^{\lambda}\theta\right]\\
 & = & \frac{-1}{\xi}(H^{a})_{\,\,\,\,\,\lambda}^{\mu}\partial^{\lambda}\square\theta.
\end{eqnarray}
From this, one also sees that the current is invariant under residual
gauge transformations of Eq.~(\ref{eq:transform}) even if $\xi\neq\infty$.
Because of the $(H^{a})_{\,\,\,\,\,\,\lambda}^{\mu}$ in front of
Eq.~(\ref{eq:btgtcurrent}), it is also invariant under the BTGT
symmetry of Eq.~(\ref{eq:BTGTsymhere}).

The BTGT current Eq.~(\ref{eq:btgtcurrent}) can be interpreted in
terms of $A_{\mu}$ by noting that the classical relationship 
\begin{equation}
A_{\mu}=\sum_{b}(H^{b})_{\,\,\,\,\,\,\mu}^{\alpha}\partial_{\alpha}\theta^{b}
\end{equation}
(see \cite{Chung:2016lhv} for details) allows us to identify
\begin{equation}
\sum_{a}\mathcal{B}_{a}^{\mu}=\partial^{\mu}\partial\cdot A(1-\frac{1}{\xi})-\square A^{\mu}+J_{U(1)}^{\mu}\label{eq:sum-of-currents}
\end{equation}
where

\begin{equation}
J_{U(1)}^{\mu}\equiv-\left[ie\phi^{*}\partial^{\mu}\phi+h.c.\right]-2e^{2}|\phi|^{2}A^{\mu}.
\end{equation}
Hence, the sum of the BTGT currents itself is the equation of motion
in terms of the usual gauge field $A_{\mu}.$ As we will see below,
the equation of motion for $\theta^{a}$ differs from the equation
of motion for $A_{\mu}$ by a derivative. Hence, BTGT current can
be interpreted as the decomposition of $A^{\mu}$ equation of motion
in basis tensor components (to be distinguished from the decomposition
of $\theta^{a}$ equation of motion in basis tensor components).

To check $\mathcal{B}_{a}^{\mu}$ conservation using the classical
equations of motion, note Eq.~(\ref{eq:conservationbtgtcurrent})
can be rewritten as 
\begin{equation}
\partial\star_{a}\left[\partial\sum_{b}\partial\star_{b}\partial\theta^{b}(1-\frac{1}{\xi})-\partial\square\theta^{a}+\mathcal{C}_{U(1)a}(x)\right]=0
\end{equation}
for each $a$ choice. Since this is precisely the equation of motion
of $\theta^{a}$, the BTGT classical current is manifestly conserved
for classical $\theta^{b}$ fields satisfying the equation of motion.
On the other hand, the equation of motion for $A^{\mu}$ is the sum
of the currents shown in Eq.~(\ref{eq:sum-of-currents}) and not
the equation of motion for $\theta^{b}$ itself: i.e.~the difference
is a particular type of derivative.

Now, let's combine 
\begin{equation}
\partial_{\mu}\sum_{a}\mathcal{B}_{a}^{\mu}=0
\end{equation}
with the Noether current conservation coming from the global subgroup
of the $U(1)$ gauge symmetry (which obviously is not broken even
for finite $\xi$):
\begin{equation}
\partial_{\mu}J_{U(1)}^{\mu}=0.
\end{equation}
This means that the ordinary $U(1)$ current conservation and the
\emph{sum} of the BTGT current conservation together enforces
\begin{equation}
\frac{1}{\xi}\square\partial\cdot A=0\label{eq:sumconserve}
\end{equation}
which is the same as Eq.~(\ref{eq:residual-gauge-current}) enforced
by the residual gauge symmetry.

\section{\label{sec:A-brief-review}A brief review of effective action generating
ward identities}

We would now like to study the quantum current conservation associated
with the BTGT symmetry: i.e. generate Ward identities. For this goal,
we briefly review here the effective action formalism for generating
Ward identities \cite{Ward:1950xp}. This allows us to then reduce
the generation of Ward identities to a set of functional derivatives,
which we will use to obtain explicit Ward identities for the BTGT
and gauge symmetry.

Let $\varphi$ be a vector of fields (e.g. $\varphi=(\theta^{a},\phi,\phi^{*},...)$
and $J=(J_{a},J_{\phi},J_{\phi^{*}},...)$). The generating functional
$W$ for the connected Green's functions $\mathcal{G}_{c}^{(n)}$
is
\begin{eqnarray}
iW[J] & \equiv & \ln Z[J]\\
 & = & \sum_{n=0}^{\infty}\frac{i^{n}}{n!}\int d^{4}x_{1}...d^{4}x_{n}J(x_{1})...J(x_{n})\mathcal{G}_{c}^{(n)}(x_{1},...,x_{n})
\end{eqnarray}
and the effective action is its Legendre transform:
\begin{equation}
\Gamma[\bar{\varphi}]\equiv W[J(\bar{\varphi})]-\int d^{4}xJ(\bar{\varphi})\bar{\varphi}\label{eq:gammaaslegendre}
\end{equation}
where
\begin{equation}
\bar{\varphi}(x)\equiv\frac{\delta W[J]}{\delta J(x)}.\label{eq:classical}
\end{equation}
The effective action $\Gamma$ can be interpreted as a collection
of amplitudes
\begin{equation}
\Gamma[\varphi_{c}]=\sum_{n}\frac{1}{n!}\int d^{4}x_{1}...d^{4}x_{n}\varphi_{c}(x_{1})...\varphi_{c}(x_{n})\Gamma^{(n)}(x_{1},...,x_{n})
\end{equation}
where $\Gamma^{(n)}(x_{1},...,x_{n})$ are external momenta truncated
(with full propagators) 1PI graphs. 

It is well known that one can choose counter terms perturbatively
such that only gauge invariant counter terms renormalize the theory
as long as the regulators do not spoil gauge invariance. This leads
to the gauge invariant effective action if we subtract out the tree-level
gauge-fixing action from the total effective action:
\begin{equation}
\hat{\Gamma}[\varphi_{c}]\equiv\Gamma-S_{GF}.
\end{equation}
At this point, one simply applies the functional derivative representation
of the symmetry transformation and pick out functionally independent
coefficients to generate the Ward identities.

As a simple illustration, consider scalar QED in the usual $A_{\mu}$
gauge field representation. The gauge transformation 
\begin{equation}
\delta\phi=ie\theta\phi,\,\,\,\,\,\,\,\delta\phi^{*}=-ie\theta\phi^{*},\,\,\,\,\,\delta A_{\mu}=\partial_{\mu}\theta\label{eq:ordinaryqed-gauge-transform}
\end{equation}
mixes the 3-point function involving $(A_{\mu},\phi^{*},\phi)$ fields
and 2-point functions obtained from removing $A_{\mu}$ due to the
inhomogeneous nature of $\delta A_{\mu}$. Hence, the relevant components
of the effective action are
\begin{eqnarray}
\hat{\Gamma} & \ni & I\equiv\int d^{4}xd^{4}yd^{4}z\phi^{*}(x)\phi(y)A^{\mu}(z)\hat{\Gamma}_{\mu}^{(1,1,1)}(x,y,z)+\nonumber \\
 &  & \int d^{4}xd^{4}y\phi^{*}(x)\phi(y)\hat{\Gamma}^{(1,1,0)}(x,y).
\end{eqnarray}
To apply the symmetry transformations on this object, one needs a
functional representation of Eq.~(\ref{eq:ordinaryqed-gauge-transform}):
\begin{equation}
\mathcal{O}_{0}(\phi^{*},\phi,A^{\nu})=(-ie\theta\phi^{*},ie\theta\phi,\partial^{\nu}\theta)
\end{equation}
where
\begin{equation}
\mathcal{O}_{0}\equiv\int d^{4}y'\partial^{\nu}\theta(y')\frac{\delta}{\delta A^{\nu}(y')}+ie\int d^{4}y'\theta(y')\phi(y')\frac{\delta}{\delta\phi(y')}-ie\int d^{4}y'\theta(y')\phi^{*}(y')\frac{\delta}{\delta\phi^{*}(y')}.\label{eq:functionalu1orig}
\end{equation}
Ward identities are generated through
\begin{equation}
\frac{\delta}{\delta\theta(z_{3})}\frac{\delta}{\delta\phi^{*}(z_{2})}\frac{\delta}{\delta\phi(z_{1})}\mathcal{O}_{0}I|_{A=0}=0.
\end{equation}
Explicitly, one finds the well known Ward identity:
\begin{equation}
\partial^{\mu}\hat{\Gamma}_{\mu}^{(1,1,1)}(z_{2},z_{1},z_{3})=ie\delta^{(4)}(z_{1}-z_{3})\hat{\Gamma}^{(1,1,0)}(z_{2},z_{1})-ie\delta^{(4)}(z_{2}-z_{3})\hat{\Gamma}^{(1,1,0)}(z_{2},z_{1})
\end{equation}

\section{\label{sec:BTGT-formalism-Ward}BTGT formalism Ward identities}

In this section, we use the method of the previous section to derive
the Ward identities in the BTGT formalism both for the $U(1)$ gauge
symmetry and the BTGT symmetry.

\subsection{$U(1)$ in BTGT formalism}

Here we consider the $U(1)$ gauge symmetry transformations of Eq.~(\ref{eq:ordinarygaugeu1inbtgt}).
The functional derivative representation analog of Eq.~(\ref{eq:functionalu1orig})
for this symmetry is

\begin{equation}
\mathcal{O}_{U(1)}\equiv\sum_{f}\int d^{4}y'\theta(y')\frac{\delta}{\delta\theta^{f}(y')}+ie\left[\int d^{4}y'\theta(y')\phi(y')\frac{\delta}{\delta\phi(y')}-\int d^{4}y'\theta(y')\phi^{*}(y')\frac{\delta}{\delta\phi^{*}(y')}\right].
\end{equation}

First, applying this operator to the 2-point function
\begin{equation}
\hat{\Gamma}\ni\frac{1}{2}\int d^{4}xd^{4}y\sum_{ab}\hat{\Gamma}_{ab}^{(0,0,2)}(x,y)\theta^{a}(x)\theta^{b}(y),
\end{equation}
we find
\begin{equation}
I_{2}\equiv\sum_{ab}\int d^{4}x\int d^{4}y\hat{\Gamma}_{ab}^{(0,0,2)}(x,y)\theta(x)\theta^{b}(y)+\sum_{ab}\int d^{4}y\int d^{4}x\hat{\Gamma}_{ab}^{(0,0,2)}(x,y)\theta^{a}(x)\theta(y)=0
\end{equation}
Next, we apply the functional independence constraint to obtain
\begin{equation}
\frac{\delta}{\delta\theta(z_{1})}\frac{\delta}{\delta\theta^{f}(z_{2})}I_{2}=\sum_{a}\left(\hat{\Gamma}_{af}^{(0,0,2)}(z_{1},z_{2})+\int d^{4}x\hat{\Gamma}_{fa}^{(0,0,2)}(z_{2},z_{1})\right)=0
\end{equation}
Since the 2-point function is symmetric in $af$ symbol, we conclude
\begin{equation}
\sum_{a}\hat{\Gamma}_{ab}^{(0,0,2)}(z,x)=0\label{eq:U1wardidentity2pt}
\end{equation}
and
\begin{equation}
\sum_{b}\hat{\Gamma}_{ab}^{(0,0,2)}(x,z)=0.
\end{equation}
(Since we have been a bit overexplicit here, we will henceforth omit
writing redundant combinations derivable from the index/variable permutations.)

Let's see if this is satisfied by the tree-level propagator. Note
that in Fourier space, the 2-point coefficient is given by the inverse
of the propagator:
\begin{equation}
\Gamma_{fc}^{(0,0,2)}(k)=i(\Delta^{-1})_{fc}
\end{equation}
where $\Delta_{bc}(k)$ is the propagator defined by
\begin{equation}
\langle\theta^{b}(y)\theta^{c}(z)\rangle=\int\frac{d^{4}l}{(2\pi)^{4}}e^{-il\cdot(y-z)}\Delta_{bc}(l).
\end{equation}
To obtain $\hat{\Gamma}_{ab}$, we subtract out the gauge fixing term:
\begin{equation}
S_{GF}=\frac{-1}{2\xi}\int d^{4}x\theta^{a}(H^{a})_{\,\,\,\,\,\mu}^{\alpha}(H^{b})_{\,\,\,\,\,\beta}^{\lambda}\partial^{\mu}\partial_{\alpha}\partial^{\beta}\partial_{\lambda}\theta^{b}.
\end{equation}
We can rewrite this as 
\begin{equation}
S_{GF}=\frac{1}{2!}\int d^{4}xd^{4}z\Gamma_{ab}^{GF}(x,z)\theta^{a}(x)\theta^{b}(z)
\end{equation}
and Fourier transform as
\begin{equation}
\Gamma_{ab}^{GF}(x,z)\equiv\int\frac{d^{4}k}{(2\pi)^{4}}e^{ik\cdot(x-z)}\Gamma_{ab}^{GF}(k)\label{eq:Gammakdef}
\end{equation}
to obtain
\begin{equation}
\Gamma_{ab}^{GF}(k)=\frac{-1}{\xi}(H^{a})_{\,\,\,\,\,\mu}^{\alpha}(H^{b})_{\,\,\,\,\,\beta}^{\lambda}k^{\mu}k_{\alpha}k^{\beta}k_{\lambda}.
\end{equation}
The gauge fixing term subtraction therefore gives
\begin{equation}
\hat{\Gamma}_{fc}^{(0,0,2)}(k)=\Gamma_{fc}^{(0,0,2)}(k)+\frac{1}{\xi}(k\star_{f}k)(k\star_{c}k)
\end{equation}
and the Ward identity is
\begin{equation}
\boxed{\sum_{f}\left[i(\Delta^{-1})_{fc}+\frac{1}{\xi}(k\star_{f}k)(k\star_{c}k)\right]=0}\label{eq:momspace-2pt-ward}
\end{equation}
to all orders in perturbation theory.

At tree level, we have after subtraction the expression
\begin{equation}
\hat{\Gamma}_{fc}^{(0,0,2)}(k)=-k^{2}\delta_{\,\,\,\, c}^{f}k\star_{f}k+(k\star_{f}k)(k\star_{c}k).
\end{equation}
We can sum over $f$ and find
\begin{equation}
\sum_{f}\hat{\Gamma}_{fc}^{(0,0,2)}(k)=-k^{2}k\star_{c}k+k^{2}(k\star_{c}k)=0
\end{equation}
confirming the Ward identity for $U(1)$ at tree-level.

Next, let's consider the 3-point function. Operating $\mathcal{O}_{U(1)}$
on 
\begin{equation}
\hat{\Gamma}\ni\sum_{c}\int d^{4}xd^{4}yd^{4}z\phi^{*}(x)\phi(y)\theta^{c}(z)\hat{\Gamma}_{c}^{(1,1,1)}(x,y,z)+\int d^{4}xd^{4}y\phi^{*}(x)\phi(y)\hat{\Gamma}^{(1,1,0)}(x,y)
\end{equation}
we find for the 3-point function Ward identity
\begin{equation}
-\sum_{c}\hat{\Gamma}_{c}^{(1,1,1)}(z_{2},z_{1},z_{3})=ie\delta^{(4)}(z_{1}-z_{3})\hat{\Gamma}^{(1,1,0)}(z_{2},z_{1})-ie\delta^{(4)}(z_{2}-z_{3})\hat{\Gamma}^{(1,1,0)}(z_{2},z_{1}).\label{eq:U1gaugeward3pt1}
\end{equation}
In Fourier space, with Feynman diagrams drawn with outgoing scalar
momentum $p_{2}=p$ and incoming scalar momentum $p_{1}=-q$, and
incoming $\theta^{c}$ field momentum $p_{3}=-k$, we find
\begin{equation}
\boxed{-\sum_{c}\hat{G}_{c}^{(1,1,1)}(p,-q,-k)=ie\left[\hat{G}^{(1,1,0)}(p,-q-k)-\hat{G}^{(1,1,0)}(p-k,-q)\right]}\label{eq:U1gaugeward}
\end{equation}
where $\hat{G}_{c}$ is the Fourier space representation of $\hat{\Gamma}_{c}$:
e.g.
\begin{equation}
\hat{\Gamma}_{c}^{(1,1,1)}(z_{2},z_{1},z_{3})=\sum_{c}\int\prod_{n=1}^{3}\frac{d^{4}k_{n}}{(2\pi)^{4}}e^{ik_{n}\cdot z_{n}}\hat{G}_{c}^{(1,1,1)}(k_{2},k_{1},k_{3}).
\end{equation}
At tree level, Eq.~(\ref{eq:U1gaugeward}) is
\begin{eqnarray}
-\sum_{c}e(q+p)\star_{c}k & = & (ie)i\left[ip\cdot(i[-q-k])-i(p-k)\cdot(i[-q])\right]\\
 & = & -e(p+q)\cdot k
\end{eqnarray}
which indeed is an identity.

\subsection{BTGT symmetry}

Next, we investigate the Ward identities associated with the BTGT
symmetry Eq.~(\ref{eq:BTGTsymhere}). The functional derivative representation
is 
\begin{equation}
\delta\theta^{a}\equiv\mathcal{O}_{B}^{(a)}\theta^{a}
\end{equation}
where
\begin{equation}
\mathcal{O}_{B}^{(a)}\equiv\int d^{4}z\utilde{Z}^{a}(z)\frac{\delta}{\delta\theta^{a}(z)}\,\,\,\,\,\,\mbox{no sum over }a
\end{equation}

Apply this to the two-point function
\begin{equation}
\Gamma\ni\frac{1}{2}\sum_{ab}\int d^{4}xd^{4}y\hat{\Gamma}_{ab}^{(0,0,2)}(x,y)\theta^{a}(x)\theta^{b}(y)
\end{equation}
which is symmetric in both $a\leftrightarrow b$ and $x\leftrightarrow y$.
We find 
\begin{equation}
\boxed{\sum_{c}\int d^{4}x\hat{\Gamma}_{cb}^{(0,0,2)}(x,y)\utilde{Z}^{c}(x)=0.}\label{eq:btgtwardidentity}
\end{equation}
We cannot take a simple functional derivative with respect to $\utilde{Z}^{c}(x)$
because that variable is constrained. One can solve the constraint
trivially in Fourier space:
\begin{equation}
\utilde{Z}^{a}(x)=\int\frac{d^{4}k}{(2\pi)^{4}}e^{i\sum_{b\neq a}k\star_{b}x}F^{a}(k)
\end{equation}
where $F^{a}(k)$ is unconstrained. Eq.~(\ref{eq:btgtwardidentity})
implies that the 2-point contribution in momentum space satisfies
\begin{equation}
\boxed{\hat{\Gamma}_{cb}(K_{\beta}^{(\perp c)})=0}\label{eq:BTGTwardidentity}
\end{equation}
where
\begin{equation}
K_{\beta}^{(\perp c)}(k)\equiv\sum_{b\neq c}k_{\alpha}(H^{b})_{\,\,\,\,\,\,\beta}^{\alpha}
\end{equation}
is the momentum vector with the $(H^{c})_{\,\,\,\,\,\beta}^{\alpha}$
component projected out.

This can be trivially checked with the tree level action:
\begin{equation}
\hat{\Gamma}_{fc}(q)=-q^{2}\delta^{fc}q\star_{f}q+(q\star_{f}q)(q\star_{c}q)
\end{equation}
where there is no sum over the repeated indices here. Since 
\begin{equation}
K^{(\perp c)}(k)\cdot K^{(\perp c)}(k)=\sum_{m\neq c}k\star_{m}k
\end{equation}
and 
\begin{equation}
K^{(\perp c)}(k)\star_{c}K^{(\perp c)}(k)=0
\end{equation}
we see that Eq.~(\ref{eq:BTGTwardidentity}) is satisfied at the
tree level.

At the 3-point function level, the relevant terms in the effective
action are
\begin{equation}
\hat{\Gamma}\ni\sum_{c}\int d^{4}xd^{4}yd^{4}z\phi^{*}(x)\phi(y)\theta^{c}(z)\hat{\Gamma}_{c}^{(1,1,1)}(x,y,z)+\int d^{4}xd^{4}y\phi^{*}(x)\phi(y)\hat{\Gamma}^{(1,1,0)}(x,y).
\end{equation}
Acting on this with $\mathcal{O}_{B}^{(a)}$, imposing BTGT symmetry,
and taking functional derivatives with respect to unconstrained variables,
we find
\begin{equation}
\boxed{\int d^{4}z\utilde{Z}^{a}(z)\hat{\Gamma}_{a}^{(1,1,1)}(x,y,z)=0}.\label{eq:3-pointbtgt}
\end{equation}
As in Eq.~(\ref{eq:BTGTwardidentity}), we can simplify this expression
further in Fourier space where the constraint on $\utilde{Z}^{a}(z)$
can be trivially solved. Defining
\begin{equation}
\hat{\Gamma}_{a}^{(1,1,1)}(x_{1},x_{2},x_{3})=\int\prod_{n=1}^{3}\frac{d^{4}k_{n}}{(2\pi)^{4}}e^{i\sum_{r=1}^{3}k_{r}\cdot x_{r}}\hat{G}_{a}^{(1,1,1)}(k_{1},k_{2},k_{3})
\end{equation}
Eq.~(\ref{eq:3-pointbtgt}) becomes
\begin{equation}
\boxed{\hat{G}_{a}^{(1,1,1)}(k_{1}^{\mu},k_{2}^{\mu},-[K^{(\perp a)}(k)]^{\mu})=0}\label{eq:3-point-btgt-fourier}
\end{equation}
Since 
\begin{equation}
\hat{G}_{a}^{(1,1,1)}(k_{1},k_{2},k_{3})\propto-e(k_{1}-k_{2})\star_{a}k_{3}
\end{equation}
we see that Eq.~(\ref{eq:3-point-btgt-fourier}) is manifestly satisfied
at tree level owing to the orthogonality property manifest in Eq.~(\ref{eq:orthogonal}).

\section{\label{sec:1-loop-renormalization}1-loop renormalization}

One way to check the quantum stability of the BTGT formalism is to
explicitly renormalize at 1-loop. Hence, in this section, we write
down the Feynman rules for the scalar QED specified by Eqs.~(\ref{eq:btgtL1})-(\ref{eq:BTGTL4})
in the BTGT formalism and explicitly renormalize at 1-loop. We will
employ dimensional regularization and check that it is consistent
with the BTGT Ward identity (and the $U(1)$ Ward identity). Minimal
subtraction renormalization scheme is used whenever we want to compare
the renormalization constants $Z_{i}$ with each other. Some of the
useful identities for the computation are given in the Appendix.

\subsection{Action and Feynman rules}

\begin{figure}
\begin{centering}
\includegraphics[scale=0.8]{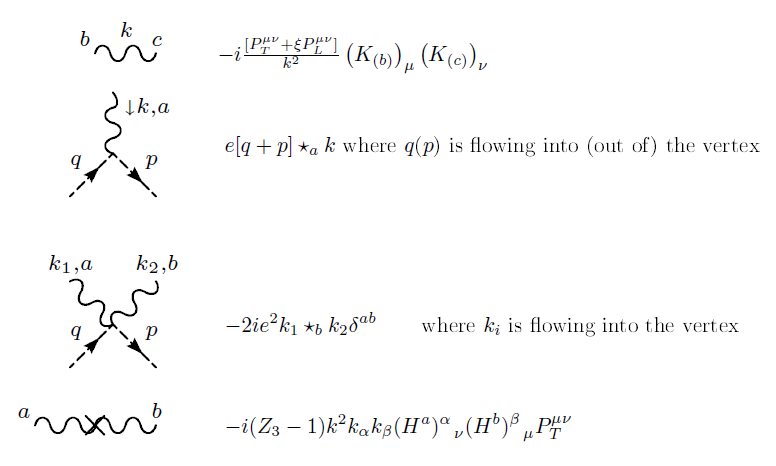}
\par\end{centering}

\protect\caption{\label{fig:Feynman-diagram-for}Feynman rules for the BTGT theory.
The counter term for the $\langle\theta^{a}\theta^{b}\rangle$ is
shown explicitly to emphasize that only gauge invariant counter terms
are needed just as in the usual connection formalism.}
\end{figure}

Here, we will write down the Feynman rules for scalar QED in the BTGT
formulation specified by Eqs.~(\ref{eq:btgtL1})-(\ref{eq:BTGTL4}).
We will use the summation convention that we sum over all repeated
Latin indices unless specified otherwise. For our objective of testing
the BTGT coupling at one loop, we have set the scalar quartic self-coupling
to zero at tree level. The Feynman rules for this scalar QED theory
is shown in Fig.~\ref{fig:Feynman-diagram-for}.   In the figure, the photon propagator has a factor written in terms of
\begin{equation}
 \left(K_{(b)}\right)_{\mu}\equiv(H^{b})_{\,\,\,\,\,\mu}^{\psi}\frac{k_{\psi}}{k_{\alpha}k_{\beta}(H^{b})^{\alpha\beta}}=(H^{b})_{\,\,\,\,\,\mu}^{\psi}\frac{k_{\psi}}{k \star_b k}.
\end{equation}
In the rest of this
section, we use these Feynman rules in the Feynman gauge ($\xi=1$)
for an explicit 1-loop renormalization of scalar QED.

\subsection{Vacuum Polarization}

\begin{figure}
\begin{centering}
\includegraphics{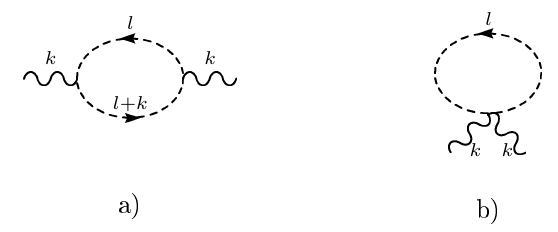}
\par\end{centering}

\protect\caption{\label{fig:Vacuum-polarization}Vacuum polarization diagrams.}
\end{figure}

In this subsection, we compute the two vacuum polarization graphs
shown in Fig.~\ref{fig:Vacuum-polarization}. All diagrams in this
subsection and the subsequent subsections will be computed in the
Feynman gauge.

The first graph we consider is the one of Fig.~\ref{fig:Vacuum-polarization}a).
The Feynman rules give
\begin{equation}
i\Pi_{1}^{ab}=\int\frac{d^{4}l}{(2\pi)^{4}}e(2l+k)\star_{a}k\left(\frac{i}{l^{2}-m^{2}}\right)\left[{\color{red}-}e(2l+k)\star_{b}k\right]\left(\frac{i}{(k+l)^{2}-m^{2}}\right).
\end{equation}
The structure of this diagram is similar to that of ordinary scalar
QED other than the appearance of star products in the numerator:
\begin{equation}
i\Pi_{1}^{ab}=k_{\alpha}(H^{a})^{\alpha\mu}k_{\beta}(H^{b})^{\beta\nu}(I_{1}^{{\rm ordinary}})_{\mu\nu}\label{eq:Pi1}
\end{equation}
where 
\begin{equation}
(I_{1}^{{\rm ordinary}})_{\mu\nu}\equiv2e^{2}\left[g_{\mu\nu}\left(-\frac{1}{6}k^{2}+m^{2}\right)+\frac{1}{6}k_{\mu}k_{\nu}\right]\frac{i}{(4\pi)^{2}}\left(\frac{2}{\epsilon}+...\right).
\end{equation}
The quartic vertex graph of Fig.~\ref{fig:Vacuum-polarization}b)
gives
\begin{eqnarray}
i\Pi_{2}^{ab} & = & ({\color{red}-}1)\left(-2ie^{2}k\star_{b}k\delta^{ab}\right)\int\frac{d^{4}l}{(2\pi)^{4}}\frac{i}{l^{2}-m^{2}}\\
 & = & k\star_{b}k\delta^{ab}I_{2}^{{\rm ordinary}}\label{eq:Pi2}
\end{eqnarray}
where 
\begin{equation}
I_{2}^{{\rm ordinary}}=-(2e^{2})\frac{i}{(4\pi)^{2}}\frac{2}{\epsilon}m^{2}+...
\end{equation}
We can add Eqs.~(\ref{eq:Pi1}) and $(\ref{eq:Pi2})$ to obtain
\begin{equation}
i\Pi_{1}^{ab}(k)+i\Pi_{2}^{ab}(k)=+k_{\alpha}(H^{a})^{\alpha\mu}k_{\beta}(H^{b})^{\beta\nu}2e^{2}\left(-\frac{1}{6}k^{2}\right)P_{T\mu\nu}\frac{i}{(4\pi)^{2}}\left(\frac{2}{\epsilon}+...\right).\label{eq:pis}
\end{equation}
Note that the scalar mass dependent longitudinal mode has decoupled
from the photon propagator, restoring gauge invariance (i.e.~leaving
the term proportional to the transverse projection operator $P_{T\mu\nu}$).
Eq.~(\ref{eq:pis}) generates the counter term
\begin{equation}
Z_{3}=1-\frac{e^{2}}{3}\frac{1}{(4\pi)^{2}}\frac{2}{\epsilon}
\end{equation}
for example in the minimal subtraction scheme.

Let's explicitly check that the Ward identities in the BTGT formalism
are satisfied to one-loop by dimensional regularization just used.
The $U(1)$ Ward identity Eq.~(\ref{eq:momspace-2pt-ward}) is satisfied
by Eq.~(\ref{eq:pis}) since 
\begin{eqnarray}
\sum_{a}\left(i\Pi_{1}^{ab}+i\Pi_{2}^{ab}\right) & \propto & \sum_{a}k_{\alpha}(H^{a})^{\alpha\mu}P_{T\mu\nu}\frac{2}{\epsilon}...\\
 & = & 0
\end{eqnarray}
where we used Eq.~(\ref{eq:identity}). That is why we were able
to absorb the divergence using only gauge invariant counter terms.
Next, the BTGT Ward identity Eq.~(\ref{eq:BTGTwardidentity}) is
satisfied by Eq.~(\ref{eq:pis}) since
\begin{eqnarray}
i\Pi_{1}^{ab}(K_{\beta}^{(\perp a)}(k))+i\Pi_{2}^{ab}(K_{\beta}^{(\perp a)}(k)) & \propto & \sum_{j\neq a}k_{\lambda}(H^{j})_{\,\,\,\,\,\,\alpha}^{\lambda}(H^{a})^{\alpha\mu}\frac{2}{\epsilon}...\\
 & = & 0
\end{eqnarray}
where we used Eq.~(\ref{eq:orthogonal}). Hence, dimensional regularization
also preserves the 2-point function BTGT Ward identity.

\subsection{Vertex correction}

\begin{figure}
\begin{centering}
\includegraphics[scale=0.8]{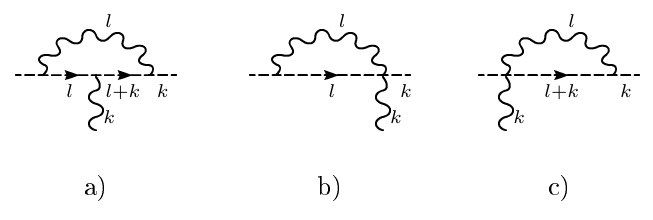}
\par\end{centering}

\protect\caption{\label{fig:One-loop-vertex-corrections}One loop cubic vertex corrections
when the scalar quartic self-coupling is zero at tree level. Note
that we have set the incoming scalar momentum to zero to simplify
the computation as is sometimes done \cite{Srednicki:2007qs}.}
\end{figure}

In this subsection, we compute the vertex corrections shown in Fig.~\ref{fig:One-loop-vertex-corrections}.
Let's first consider the diagram Fig.~\ref{fig:One-loop-vertex-corrections}a).
We find with a $\theta^{c}$ insertion
\begin{eqnarray}
i\Gamma_{c}^{(1)} & = & \sum_{a,b}\int\frac{d^{4}l}{(2\pi)^{4}}e(k_{1}+l)\star_{a}l\left(\frac{i}{l^{2}-m^{2}}\right)e(2l+k)\star_{c}k\left(\frac{i}{(l+k)^{2}-m^{2}}\right)\left[{\color{red}-}e(2k+l)\star_{b}l\right]\times\nonumber \\
 &  & \left(\frac{-i}{l^{2}}\right)\delta^{ab}\frac{1}{l\star_{b}l}\\
 & = & -ik\star_{c}\left(i\Gamma^{{\rm ordinary\,\,1}}\right)
\end{eqnarray}
where
\begin{eqnarray}
i\left(\Gamma^{{\rm ordinary\,\,1}}\right)^{\kappa} & \equiv & e^{3}\int\frac{d^{4}l}{(2\pi)^{4}}\frac{(2l+k)^{\kappa}}{l^{2}-m^{2}}\frac{(2k+l)^{\lambda}}{(l+k)^{2}-m^{2}}\frac{l_{\lambda}}{l^{2}}\\
 & = & e^{3}\left[k^{\kappa}\frac{i}{(4\pi)^{2}}\frac{2}{\epsilon}+...\right].
\end{eqnarray}
Next, the diagram Fig.~\ref{fig:One-loop-vertex-corrections}b) evaluates
to
\begin{eqnarray}
i\Gamma_{c}^{(2)} & = & \sum_{ab}\int\frac{d^{4}l}{(2\pi)^{4}}e(l)\star_{a}l\left(\frac{i}{l^{2}-m^{2}}\right)\left(-2ie^{2}[-l]\star_{b}k\delta^{cb}\right)\left(\frac{-i}{l^{2}}\delta^{ab}\frac{1}{l\star_{b}l}\right)\\
 & = & 2ie^{3}k_{\gamma}(H^{c})^{\lambda\gamma}\int\frac{d^{4}l}{(2\pi)^{4}}\frac{1}{l^{2}-m^{2}}\frac{l_{\lambda}}{l^{2}}
\end{eqnarray}
which vanishes because of Lorentz symmetry and the fact that we have
set the incoming scalar momentum to zero.

The most interesting diagram is Fig.~\ref{fig:One-loop-vertex-corrections}c)

\begin{eqnarray}
i\Gamma_{c}^{(3)} & = & \sum_{ab}\int\frac{d^{4}l}{(2\pi)^{4}}\left(-2ie^{2}{\color{red}k}{\color{red}\star_{a}l}\delta^{ca}\right)\frac{i}{(l+k)^{2}-m^{2}}\left(e[l+2k]\star_{b}[-{\color{red}l}]\right)\left(\frac{-i}{l^{2}}\delta^{ab}\frac{1}{{\color{red}l\star_{b}l}}\right)\\
 & = & -i{\color{red}k_{\gamma}(H^{c})^{\gamma\kappa}}(I_{\xi_{3}})_{\kappa}+i\underline{\Gamma}_{c}^{(3)}\label{eq:diagramcsplit}
\end{eqnarray}
where
\begin{eqnarray}
(I_{\xi_{3}})_{\kappa} & \equiv & -2e^{3}\int\frac{d^{4}l}{(2\pi)^{4}}\frac{1}{(l+k)^{2}-m^{2}}\frac{l_{\kappa}}{l^{2}}\\
 & = & e^{3}k_{\kappa}\left(\frac{i}{(4\pi)^{2}}\frac{2}{\epsilon}+...\right).
\end{eqnarray}
and
\begin{equation}
i\underline{\Gamma}_{c}^{(3)}\equiv\boxed{4ie^{3}{\color{red}k_{\alpha}(H^{c})^{\alpha\psi}}k_{\beta}{\color{red}(H^{c})^{\beta\delta}}\int\frac{d^{4}l}{(2\pi)^{4}}\frac{{\color{red}l_{\psi}}{\color{red}l_{\delta}}}{(l+k)^{2}-m^{2}}\frac{1}{l^{2}}\frac{1}{{\color{red}l\star_{c}l}}}.
\end{equation}
which a priori looks different from the usual computation particularly
because of the tensor structure in the denominator. By boosting to
the diagonal frame of $H^{c}$, we can simplify this. Afterwards,
we boost back to find
\begin{equation}
i\underline{\Gamma}_{c}^{(3)}=4ie^{3}(H^{c})^{\mu\nu}k_{\mu}k_{\nu}\left(\frac{i}{(4\pi)^{2}}\left[\frac{2}{\epsilon}+...\right]\right).
\end{equation}

Combining this with Eq.~(\ref{eq:diagramcsplit}), we find the third
diagram contributes
\begin{equation}
i\Gamma_{c}^{(3)}(k)=-3e^{3}\frac{k\star_{c}k}{(4\pi)^{2}}\left[\frac{2}{\epsilon}+...\right].
\end{equation}
Combining the three vertex correction diagrams, we thus arrive at
\begin{eqnarray}
i\Gamma_{c} & = & i\sum_{n=1}^{3}\Gamma_{c}^{(n)}\\
 & = & -ik\star_{c}\left(i\Gamma^{{\rm ordinary}}\right)\label{eq:3-point}
\end{eqnarray}
where
\begin{eqnarray}
i\left(\Gamma^{{\rm ordinary}}\right)^{\kappa} & \equiv & -i\frac{3e^{3}}{8\pi^{2}\epsilon}k^{\kappa}+k^{\kappa}\frac{i}{(4\pi)^{2}}\frac{2}{\epsilon}\xi e^{3}|_{\xi=1}+{\rm finite}\\
 & = & -i\frac{e^{3}}{4\pi^{2}\epsilon}k^{\kappa}+{\rm finite}
\end{eqnarray}
This leads to (for example in the minimal subtraction renormalization
scheme) 
\begin{equation}
Z_{1}=1+\frac{e^{2}}{4\pi^{2}\epsilon}.
\end{equation}

With Eq.~(\ref{eq:3-point}), we can also check that the 3-point
BTGT Ward identity Eq.~(\ref{eq:3-point-btgt-fourier}) is preserved
by the dimensional regularization. Before adding the counter term,
we have the regularized contribution 
\begin{eqnarray}
\hat{G}_{a}^{(1,1,1)}(k_{1}^{\mu},k_{2}^{\mu},-[K^{(\perp a)}(k)]^{\mu}) & \ni & -k\star_{a}k\frac{e^{3}}{4\pi^{2}\epsilon}|_{k=-K^{(\perp a)}(k)}\\
 & = & 0
\end{eqnarray}
satisfying the BTGT Ward identity. To check the $U(1)$ Ward identity,
we still need the 2-point function for the scalars.

\subsection{Scalar kinetic correction}

\begin{figure}
\begin{centering}
\includegraphics{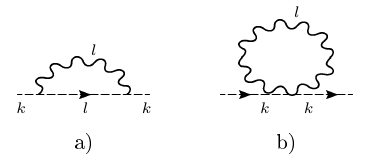}
\par\end{centering}

\protect\caption{\label{fig:One-loop-scalarmass-correction}One loop scalar kinetic
corrections when the scalar quartic self-coupling is zero at tree
level. }
\end{figure}

Let's now compute the 1-loop scalar kinetic correction to check the
Ward identity involving the 3-point and 2-point functions. Fig.~\ref{fig:One-loop-scalarmass-correction}a)
gives 
\begin{equation}
i\Pi_{(s1)}=\sum_{ab}\int\frac{d^{4}l}{(2\pi)^{4}}e[2k+l]\star_{a}l\left(\frac{i}{(k+l)^{2}-m^{2}}\right)e[2k+l]\star_{b}[-l]\left(\frac{-i}{l^{2}}\delta^{ab}\frac{1}{l\star_{b}l}\right)
\end{equation}
where the novel tensor structure in the denominator can be handled
by boosting to to the $(H^{b})_{\,\,\,\,\,\nu}^{\mu}$ diagonal frame
as in the vertex corrections. This results in 
\begin{equation}
i\Pi_{(s1)}=-2e^{2}k^{2}\frac{i}{(4\pi)^{2}}\frac{2}{\epsilon}+...\label{eq:firstmasscorrresult}
\end{equation}
 which matches the usual computation results. The diagram of Fig.~\ref{fig:One-loop-scalarmass-correction}b)
gives
\begin{eqnarray}
i\Pi_{(s2)} & = & -\sum_{ab}\int\frac{d^{4}l}{(2\pi)^{4}}2ie^{2}l\star_{b}l\delta^{ab}\frac{-i}{l^{2}}\frac{1}{l\star_{b}l}+...\\
 & \propto & \int\frac{d^{4}l}{(2\pi)^{4}}\frac{1}{l^{2}}+....
\end{eqnarray}
which has no novel tensor structure (as the numerator and the denominator
cancel) and does not contribute to the $Z_{2}$ counter term in dimensional
regularization as usual. 

Using Eq.~(\ref{eq:firstmasscorrresult}) and $i(Z_{2}-1)k^{2}$
counter term, we thus find
\begin{equation}
Z_{2}=1+\frac{e^{2}}{4\pi^{2}\epsilon}.
\end{equation}
We explicitly see that $Z_{1}=Z_{2}$ in the minimal subtraction scheme
which is the prediction of the $U(1)$ Ward identity. 

Now, let's check that dimensional regularization satisfies the $U(1)$
Ward identity mixing the 3-point function and the 2-point function
within the BTGT formalism. Using Eq.~(\ref{eq:3-point}), we find
the left hand side of Eq.~(\ref{eq:U1gaugeward}) before adding the
counter term is
\begin{equation}
\sum_{c}\left(\frac{e^{3}}{4\pi^{2}\epsilon}k\star_{c}k+...\right)=\frac{e^{3}}{4\pi^{2}\epsilon}k\cdot k+...
\end{equation}
while we find the right hand side from Eq.~(\ref{eq:firstmasscorrresult})
to be
\begin{eqnarray}
ie\left[\hat{G}^{(1,1,0)}(k,-k)-\hat{G}^{(1,1,0)}(0,0)\right] & = & ie\left[-4e^{2}k^{2}\frac{i}{(4\pi)^{2}\epsilon}\right]\\
 & = & k^{2}\frac{e^{3}}{4\pi^{2}\epsilon}.
\end{eqnarray}
Hence, the $U(1)$ Ward identity is preserved by the dimensional regularization
within the BTGT formalism as expected.

\subsection{4-point function}

\begin{figure}
\begin{centering}
\includegraphics{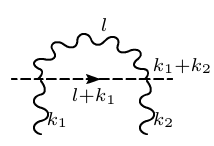}
\par\end{centering}

\protect\caption{\label{fig:4-pt-diagram}4-point function diagram that will contribute
to the $Z_{4}$ at 1-loop if we turn off the quartic scalar self-coupling
at tree level.}
\end{figure}

In this subsection we would like to check the $U(1)$ Ward identity
prediction $Z_{1}=Z_{4}$ for scalar QED. If we turn off the scalar
quartic self-interaction, we only need to evaluate the diagram shown
in Fig.~\ref{fig:4-pt-diagram}:

\begin{equation}
iV_{ab}(A)=\sum_{ce}\int\frac{d^{4}l}{(2\pi)^{4}}\left(-2ie^{2}k_{1}\star_{a}l\delta^{ac}\right)\left(\frac{i}{(l+k_{1})^{2}-m^{2}}\right)\left(-2ie^{2}k_{2}\star_{b}[-l]\delta^{be}\right)\left(\frac{-i}{l^{2}}\delta^{ce}\frac{1}{l\star_{e}l}\right)
\end{equation}
where $k_{1}$ is the incoming left external photon momentum and $k_{2}$
is the incoming right external photon momentum. Note that unlike in
the case of $A_{\mu}$ field theory, we cannot set $k_{1}=0$ to obtain
the desired counterterm. To handle the BTGT specific tensor structure
in the denominator, we can go to the $(H^{f})_{\,\,\,\,\,\nu}^{\mu}$
diagonal basis as done before to evaluate this integral. We find 
\begin{equation}
iV_{ab}(A)=4e^{4}\delta^{ab}\left(k_{1}\star_{b}k_{2}\right)\left(\frac{i}{(4\pi)^{2}}\frac{2}{\epsilon}+{\rm finite}\right)
\end{equation}
where there is no sum over the repeated indices here. Adding this
to the counter-term, the renormalization constant $Z_{4}$ can be
extracted in the minimal subtraction scheme as
\begin{equation}
Z_{4}=1+\frac{e^{2}}{4\pi^{2}\epsilon},
\end{equation}
 matching the expected result 
\begin{equation}
Z_{4}=Z_{1}=Z_{2}
\end{equation}
of the $U(1)$ Ward identity.

Given that the $Z_{i}$ at one loop has the same result as in the
ordinary formulation of scalar QED, we know that the $\beta$-function
for this theory will be the same:~i.e.~
\begin{equation}
\frac{d\alpha}{d\ln\mu}=\frac{\alpha^{2}}{6\pi}
\end{equation}
where $\alpha\equiv e^{2}(4\pi)^{-1}$. Furthermore, this explicit
computation of $Z_{4}$ was a nontrivial test of the (two $\theta$)-(two
scalar) coupling loop computation within the BTGT formalism.

\section{\label{sec:Conclusions}Conclusions}

In this work, we have investigated the Ward identities in the BTGT
formalism associated with the BTGT symmetry and the $U(1)$ symmetry.
As can be seen in Eq.~(\ref{eq:btgtcurrent}), the BTGT symmetry
current $\mathcal{B}_{a}^{\mu}$ can be classically interpreted as
the basis tensor component decomposition of $A^{\mu}$ equation of
motion. Furthermore, the $U(1)$ current conservation and the conservation
of a particular sum of the BTGT symmetry currents imply the same equation
as the residual gauge symmetry current conservation as can be seen
in Eq.~(\ref{eq:sumconserve}). The $U(1)$ Ward identities for the
two and three point functions in the BTGT formalism are displayed
in Eqs.~(\ref{eq:U1wardidentity2pt}), (\ref{eq:momspace-2pt-ward}),
(\ref{eq:U1gaugeward3pt1}), and (\ref{eq:U1gaugeward}) while those
for the BTGT Ward identities are displayed in Eqs.~(\ref{eq:btgtwardidentity}),
(\ref{eq:BTGTwardidentity}), (\ref{eq:3-pointbtgt}), and (\ref{eq:3-point-btgt-fourier}).

To check whether or not dimensional regularization is consistent with
the BTGT symmetry Ward identity and the BTGT formalism in general,
explicit one loop renormalization of scalar QED was carried out. All
two and three point Ward identities associated with BTGT symmetry
and $U(1)$ are shown to be consistent with dimensional regularization.
Novel dot products in the form of $A\star_{c}B$ appear both in the
numerator and the denominator, but the renormalization constants in
the minimal subtraction scheme are identical to the results from the
standard computational formalism. It is clear from these explicit
computations that the BTGT formalism is stable at one loop.

There are many future research directions for BTGT. It would be interesting
to see if the non-Abelian gauge theories can be expressed in the BTGT
formalism. This involves constructing a solution to the nonlinear
constraint equation in a fashion similar to what was done for the
Abelian theory. It would also be interesting to find practical applications
for this theory in computing non-local correlators or in lattice gauge
theory. This formalism should also be tested in the contexts of spontaneous
symmetry breaking and curved spacetime. Since $\theta^{a}$ field
is very close to a Wilson line and since it is charged under a BTGT
symmetry, the Abelian BTGT symmetry may be related to generalized
global symmetries \cite{Gaiotto:2014kfa}. It would be worth investigating
the precise connection. For beyond the standard model physics, it
would be interesting to see if the basis tensor fields
\begin{equation}
G_{\,\,\,\,\,\,\mu}^{\beta}(x)=\left[e^{i\theta^{a}H^{a}}\right]_{\,\,\,\,\,\,\mu}^{\beta}
\end{equation}
can be embedded into a spontaneously broken theory since these fields
are suggestive of a sigma model.
\begin{acknowledgments}
This work was supported in part by the DOE through grant DE-SC0017647. 
\end{acknowledgments}
\appendix

\section{\label{sec:Useful-identities}Useful identities}

The basis tensor matrices $(H^{a})_{\,\,\,\,\,\,\nu}^{\mu}$ were
introduced in \cite{Chung:2016lhv} to solve the non-linear constraint
equations. In the same sense in which $\gamma^{\mu}$ can be viewed
to transform under Lorentz transformations, $(H^{a})_{\,\,\,\,\,\,\,\nu}^{\mu}$
can be thought of as an object transforming as a $(1,1)$ tensor under
Lorentz transformations.

The basis tensor matrices has the following properties which are useful
for Feynman diagram computations:
\begin{equation}
\sum_{a}(H^{a})_{\,\,\,\,\,\,\nu}^{\mu}=\delta_{\,\,\,\,\,\,\,\nu}^{\mu}\label{eq:identity}
\end{equation}

\begin{equation}
(H^{a})^{\mu\nu}=(H^{a})^{\nu\mu}
\end{equation}
\begin{equation}
(H^{a})_{\,\,\,\,\,\,\nu}^{\mu}(H^{b})^{\nu\lambda}=\delta^{ab}(H^{a})^{\mu\lambda}\,\,\,\,\,\mbox{no sum over }a\label{eq:orthogonal}
\end{equation}
These matrices also serve as a kind of metric in BTGT characteristic
dot products:
\begin{equation}
A\star_{b}B\equiv A^{\mu}B^{\nu}(H^{b})_{\mu\nu}
\end{equation}
where $A$ and $B$ are Lorentz 4-vector quantities.

Some of our other conventions used in this paper are as follows:
\begin{equation}
P_{T\mu\nu}\equiv\eta_{\mu\nu}-\frac{k_{\mu}k_{\nu}}{k^{2}}
\end{equation}
\begin{equation}
P_{L\mu\nu}\equiv\frac{k_{\mu}k_{\nu}}{k^{2}}
\end{equation}
\begin{equation}
\eta_{\mu\nu}={\rm diagonal}(1,-1,-1,-1).
\end{equation}

Finally, one should note that there is a typo in equation 36 of \cite{Chung:2016lhv}.
For equation 36 to be consistent with equation 39, we should define
equation 36 to be 
\begin{equation}
(H^{a})_{\,\,\,\,\,\,\nu}^{\mu}=\psi_{(a)}^{\mu}\psi_{(a)\nu}\eta^{aa}\mbox{ \,\,\,\,\,\,\,\,\ no sum over }a
\end{equation}
where 
\begin{equation}
\psi_{(a)}^{\mu}\equiv\Lambda_{\,\,\,\,\, a}^{\mu}
\end{equation}
is the Lorentz boost matrix that boosts away from the diagonal basis.

\bibliographystyle{JHEP2}
\bibliography{btgtpaper}

\end{document}